\documentstyle[12pt]{article}
\def\intq#1{\int \frac{d^4#1}{(2\pi)^4}}
\begin{document}

\title{Variational Methods for Nuclear Systems \\ with Dynamical Mesons }
\author{${}^\dagger$ R.Cenni and
${}^\ddagger$ S.Fantoni\\~\\
${}^\dagger$ Dipartimento di Fisica dell'Universit\`a di Genova \\
Istituto Nazionale Fisica Nucleare -- Sez. di Genova \\
Via Dodecaneso 33 -- 16146 -- Genova -- Italy\\
${}^\ddagger$  Interdisciplinary Laboratory, SISSA \\
Via Beirut 2, I-34014, Trieste, Italy\\
Istituto Nazionale Fisica Nucleare, sezione di Trieste}
\date{}
\maketitle
\begin{abstract}
We derive a model Hamiltonian whose
ground state expectation value of any two-body operator coincides
with that obtained with the Jastrow correlated wave function
of the many-body Fermi
system. Using this Hamiltonian we show that the variational principle
can be extended to treat systems with dynamical mesons,
even if in this case the concept of wave function looses its meaning.
\end{abstract}

\newpage

Variational methods, based on the Fermi Hyper Netted Chain (FHNC) expansion
\cite{FHNC1,FHNC2,FHNC3}, also connected with the Correlated Basis Function
perturbation theory\cite{CBF1,CBF2},
have in the last years provided a satisfactory description of the
static properties and of the response functions of the
nuclear matter\cite{CBF3} and of other quantum fluids like, e.g.,
the liquid $^3He$.

The FHNC approach is however seemingly constrained to non
relativistic potential
theories, where the wave function concept makes sense and
correlated wave functions of the type:
\begin{equation}
\psi_G = \hat{G}\phi(r_1,\dots,r_A),
\label{CF1}
\end{equation}
with $\phi (r_1, \dots, r_A)$ being a Slater determinant and $\hat{G}$ being
a correlation operator, can be used. The simplest of such correlation
operators is given by the Jastrow Ansatz
\begin{equation}
\hat{G_J} = \prod_{i<j}^A f(r_{ij}).
\label{CF1bis}
\end{equation}
In this letter we show that a variational method exists that is
able to properly account for systems where the wave function concept
breaks down: this happens for instance when dealing with
dynamical meson
exchange, a process not reducible to the frame of a potential theory,
which will be the main topics of the present letter. The same
underlying idea however could be applied to other fields in solid state
physics (Ising and Hubbard models, for instance), and could provide
further improvements in the customary FHNC calculations since our
approach, as we shall see, can allow an energy-dependence in the
correlation function.

In this letter we wish to outline the general idea, namely the
translation of the variational problem from the setting of a trial wave
function to the setting of a trial Hamiltonian. Thus we are
able to overcome the
constraint represented by the need of defining a wave function,
so allowing the treatment of
dynamical meson exchanges (or similar situations). We exemplify this
treatment for the case of the Jastrow correlated wave function.

We  avoid here
technical difficulties by confining
ourselves to
nonrelativistic nucleons coupled to mesons but disregarding for instance
state-dependent correlations;
more detailed problem will be dealt with
separately in subsequent papers.

To achieve our goal, we have to solve three problems:
\begin{enumerate}
\item to embody the variational principle into a formalism able to
handle effective Hamiltonian, \cite{FHNC6,FHNC9}
\item to translate the variational wave function (we will limit ourselvs to the
Jastrow Ansatz) into a Hamiltonian formalism,
\item to find an effective Hamiltonian with nucleons only as true degrees
of freedom but exactly including the dynamical meson
exchange\cite{FHNC7,FHNC8,FHNC11}.
\end{enumerate}

These items are nicely expressed in the Feynman Path
Integral language \cite{FHNC6}.
Concerning item 1
the  variational procedure is derived
as follows. Let $H(\psi^\dagger, \psi)$ be the Hamiltonian of
a many-body system and
\begin{equation}
Z = \int{\cal D} [\psi^\dagger, \psi] e^{-\int\limits_0^\beta d \tau\,
H(\psi^\dagger, \psi)}\;, \label {CF2}
\end {equation}
the partition function.
The ground energy of the system is
\begin {equation}
E_0 = -\lim_{\beta \to \infty} {\log Z \over \beta} \label{CF3}\;.
\end{equation}
The partition function is of course defined in the euclidean world, so that
the Feynman-Ka\c{c}  measure and consequently
the integral in (\ref{CF2}) are well defined.

We introduce a weighted average on the space of the
Feynman-Ka\c{c}-integrable functions \cite{FHNC12} as
\begin{equation}
<f>_{H_0}\ =\ {{\int {\cal D} [\psi^\dagger, \psi] e^{-\int\limits_0^\beta
d\tau H_0} f(\psi^\dagger, \psi)} \over {\int {\cal D} [\psi^\dagger,
\psi] e^{- \int \limits_0^\beta d\tau H_0}}} \label {CF4}\;,
\end{equation}
where $e^{- \int\limits_0^\beta d\tau H_0}$
plays the role of a weight function. Actually $H_0$ will be interpreted as
a suitably chosen model Hamiltonian either solvable or at least easily
manageable.
It is worth stressing that $H_0$ is not necessarily the representation
in the Bargmann-Fock space of a true Hamiltonian (i.e. an energy-independent
hermitian operator bounded from below):
it is only requested to be real and positive-definite
in the euclidean world. Thus an explicit frequency dependence of
$H_0$ can be allowed.

Using (\ref{CF4}) and the inequality $<e^{-A}> \geq  e^{- <A>}$
we find
\begin{eqnarray}
Z &=& < e^{- \int \limits_0^\beta (H - H_0) d\tau} >_{ H_0}
\int {\cal D} [ \psi^\dagger, \psi] e^{- \int\limits_0^\beta d \tau
H_0}  \nonumber\\
&\geq& e^{- < \int \limits_0^\beta (H-H_0) d \tau >_{H_0}}
\int {\cal D} [ \psi^\dagger, \psi] e^{- \int \limits_0^\beta d \tau
H_0} \label {CF7}\;,
\end{eqnarray}
from which the result
\begin{equation}
E_0 \leq E_{var} = \displaystyle \lim_{\beta \to \infty}\
{{<\int \limits_0^\beta d \tau H >_{H_0}} \over \beta}
\label{CF8}\;,
\end {equation}
easily follows.

As for $H_0$, $H$ is not requested to be
hermitian or energy-independent: it only needs to be real
in the euclidean world. Thus (\ref{CF8}) applies to effective
 Hamiltonians as well, provided they are real after a Wick  rotation.

This procedure can be applied to a meson theory
provided an effective Hamiltonian is derived;
on the other hand we are free to choose a suitable $H_0$.
This makes the difference with the variational approach:
there we guess the ground state wave function,
here we guess an unperturbed Hamiltonian.
How could we force the two approaches to coincide?

Now we are ready to better formulate question 2): we look for an
operator $H_0$ such that the Jastrow wave function $F$
will be its ground state.
Presently we need something less, namely that $E_{var}$ coincides with the
expectation value of $H$ on the Jastrow wave function $\psi_J$ of eqs.
(\ref{CF1},\ref{CF1bis})

Let us consider the following model Hamiltonian:
\begin {equation}
H_0^{(J)} =
\Delta \sum_{k < k_F}\ a_k^\dagger a_k + {\Delta \over 2} \sum_{kpq}
\lambda(q) a^\dagger_{k+q} a_k^\dagger a_{p-q} a_p \label {CF9}\;.
\end {equation}
We claim that
\begin {equation}
E_{var}^{(J)} = {< \psi_J \vert H\vert \psi_J> \over
< \psi_J \vert \psi_J>} \label {CF10}\;,
\end{equation}
provided the correct link between $\lambda(q)$ and $f(r_{12})$ is found.

We will present here only the proof that the above
equality holds for the expectation value of the potential energy, namely
\begin {equation}
V_{var}^{(J)} = \displaystyle \lim_{\beta \to \infty}\
{{<\int \limits_0^\beta d \tau V >_{H_0}} \over \beta} =
{< \psi_J \vert V\vert \psi_J> \over
< \psi_J \vert \psi_J>} \label {CF10bis}\;.
\end{equation}
Since this result will be true independently from the particular $V$
present in the  Hamiltonian, it implies that the diagonal part of the
two-body density matrix either derived from our model Hamiltonian or
from the Jastrow Ansatz will coincide. Consequently the one-body density
matrix will do the same and the kinetic energies will also coincide.

We shall show that $V_{var}^{(J)}$ is given by  the sum of
the FHNC diagrams.

Let us expand the lhs of eq. (\ref{CF10}) in Goldstone diagrams using
the bilinear part of $H_0^{(J)}$ as unperturbed Hamiltonian. A simple
analysis shows that at each
order the quantity $V_{var}^{(J)}$ is independent from $\Delta$ (a $n^{\rm th}$
order diagram
has $n$ $\Delta\lambda(q)$ terms and $n$ energy denominators: since
each of the latter $\propto\Delta$, then all $\Delta$'s
cancel).
On the contrary the
Green's functions of the theory depends upon $\Delta$ and becomes
singular when $\Delta \to 0$.

Let us now show
the connection between the perturbative expansion of
$V_{var}^{(J)}$ and the FHNC diagrammatical expansion of
$< \psi_J\vert V \vert \psi_J >/< \psi_J\vert \psi_J >$.

We consider a Feynman diagram of the expansion of $<V>_{H_0}$
(a first-order diagram is shown in fig.1a). The solid
lines (unperturbed Green's functions) are given by
\begin{equation}
{\cal G}_0(k,k_0) = {{\theta(k-k_F)} \over {k_0 + i\eta}}
+ {{\theta (k_F - k)} \over {k_0 - \Delta - i\eta}} \label {CF11}
\end {equation}
We split ${\cal G}_0$ into two pieces:
\begin{eqnarray}
{\cal G}_0(k,k_0) &=& g_0(k,k_0) + K(k,k_0) \label{CF12}\\
g_0(k,k_0) &=& \theta(k_F - k) \left\{ {1 \over {k_0 - \Delta - i\eta}}\ -\ {1
\over {k_0 + i \eta}} \right\} \label{CF13}\\
K(k,k_0) &=& {1 \over {k_0 + i \eta}} \label {CF14}
\end {eqnarray}
Next we sort new diagrams from the Feynman graph by expanding
it by means of (\ref{CF12}). In our example three graphs are
generated (fig.1b): single lines now represent $g_0$ lines and double
lines $K$ lines.
Next we carry out the frequency integration, that will be in general
elementary if $\lambda$ is energy-independent, so coming to Goldstone
diagrams.
In each of them the $K$ lines (which are momentum-independent) cannot
leave any trace and are
shrinked to points. Tadpoles also translate into constants
and are also represented as points.
The result is shown in fig. 1c.

All the
FHNC diagrams can be reproduced in a similar way.
In fact
the frequency integration  reduces the product of the
energy denominators in a given diagram
to a factor $m \Delta^{k}$ (with $m$ integer, $k$ being the order
of the diagram), and this factor is canceled by the $\Delta^k$
coming from the interaction terms. Thus the $K$ line can only
influence the integer $m$.
The $g_0$ lines after the frequency integration become
$\theta$-functions. We remind the reader
that the Pauli correlation line of the FHNC
diagrams corresponds to the Fourier transform of the $\theta$-function, i.e.
\begin{equation}
l(k_F x) = {\nu \over \rho}\ \int {{d^3p} \over {(2\pi)^3}} \theta (k_F
- p) e^{i {\bf k} \cdot {\bf x}}\;. \label {CF15}
\end{equation}
The multiplicity factors $\nu$ coming from spin-isospin
summations necessarily coincide both in the present approach and in the FHNC
formalism, since they come from the counting of closed loops; the powers
of the density, coming from integrations over single
$g_0$ lines, also coincide with those of FHNC for dimensional
reasons.

So far we have demonstrated
the topological equivalence between the FHNC diagrams
and those of the present theory. It remains clarify the relation
between the correlation function $h=f^2-1$ of FHNC theory and the
``potential'' $\lambda$ and then to verify that the coefficients carried by
the diagrams coincide in the two expansions.

For these purposes we follow  the derivation of FHNC diagrams given in \cite
{FHNC10}.
There the two-body operator $U=2\log f^2$ is introduced and
a fictitious time dependence is assigned both to $U$
and to the potential $V$ with the results
\begin{equation}
<\Phi_0|V e^U|\Phi_0>=<\Phi_0|T\int_0^1 V(t)
e^{\int_0^1U(t)dt}|\Phi_0>\label{CF23}\;.
\end{equation}
Next, in ref.\cite{FHNC10} the r.h.s. of the above equation has been expanded
by means of the Wick theorem, and
the FHNC diagrams originally derived
in ref.\cite{FHNC1} are recovered.

Now, within the present approach,
let us consider the Feynman-Ka\c{c}   integral
\begin{equation}
\int {\cal D}[a_k(\tau),a_k^\dagger(\tau)]
e^{-\int\limits_0^\beta H_0^{(J)}(\tau)d\tau}\int_0^{\beta}V(\xi)d\xi
\label{CF24}\;,
\end{equation}
for our infinite system:
since the ``unperturbed part'' of the model Hamiltonian has the Fermi gas
Slater
determinant $|\Phi_0>$ as a ground state as in (\ref{CF23}),
we can reinterpret (\ref{CF24}) as
\begin{equation}
<\Phi_0|T\int_0^{\beta}V(\xi)d\xi
e^{-\Delta\int_0^\beta \lambda(\tau)d\tau}|\Phi_0>\;.
\end{equation}
Since the physical quantities we are interested in are
independent of $\Delta$, we can put
$\Delta={1\over \beta}$ and
make the replacement $\tau\to \tau/\beta$ to recover
(\ref{CF23}) provided we identify $U$ with $-\lambda$ in
configuration space. Next we
take the limit $\beta\to \infty$ (and consequently $\Delta\to 0$) in
order to apply (\ref{CF3}) , (\ref{CF7}) and (\ref{CF8}). Since the
value of $\Delta$ is immaterial, the equivalence between the
FHNC development and the perturbative expansion of our model Hamiltonian
is proved.
A similar proof can be done for the kinetic energy part of the Hamiltonian.

Finally, in \cite{FHNC10}, it is shown that the ladder series of fig. 2
can be summed explicitly at the prize of redefining $U$. The diagram
structure being identical, we can do the same in our case by
means of the identification
\begin{equation}
h=e^{-\lambda}-1\label{CF26}\;,
\end{equation}
which, we remind, is justified  in the limit $\Delta\to 0$.

This completes the second step of our program.
Finally, to apply the previous results to pion physicis,
we need an effective Hamiltonian
containing fermions only, but also accounting for pion exchanges.
In the context of QFT an effective action has been
derived in ref. \cite{FHNC11}.
Following the
same path we can write a lagrangian density of the form
\begin{equation}
{\cal L}={\cal L}_N+{\cal L}_\pi+i\psi^\dagger(x)\Gamma(x)\psi(x)\phi(x)
\label{CF26.1}
\end{equation}
${\cal L}_N$ and ${\cal L}_\pi$ being the free lagrangians for nucleon
and pion fields respectively and $\Gamma$ a spin-isospin operator (plus
eventually derivatives)
and then the corresponding
generating functional (the partition function in the
Minkowski world), here for simplicity with fermionic sources only:
\begin{equation}
Z[\eta,\eta^\dagger]=\int{\cal D}[\psi,\psi^\dagger,\phi]
e^{i\int dx [{\cal L}+\psi^\dagger\eta+\eta^\dagger\psi]}
\end{equation}
Since the pionic field is at most quadratic in the exponent, the
integral over $\phi(x)$ is gaussian and can be carried out explicitly,
getting
\begin{equation}
Z[\eta,\eta^\dagger]=\int{\cal D}[\psi,\psi^\dagger]e^{iS_{\rm eff}
+i\int dx(\psi^\dagger\eta+\eta^\dagger\psi)}\label{CF27}\;,
\end{equation}
where
\begin{eqnarray}
\lefteqn{S_{\rm eff}=\int dx\,dy \psi^\dagger(x) G_0^{-1}(x-y)\psi(y)}
\label{CF28}\\
&&-{1\over2}\int dx\,dy \psi^\dagger(x)\Gamma(x)\psi(x)
D_0(x-y)\psi^\dagger(y)\Gamma(y)\psi(y),\nonumber
\end{eqnarray}
$G_0$ being the free fermion propagator and $D_0$ the (fully
dynamic) pion propagator
\begin{equation}
D_0(x-y)=\intq{q}e^{iq\cdot(x-y)}{1\over q_0^2-{\bf
q}^2-m_\pi^2}\label{CF29}\;,
\end{equation}

The last step is to apply eq. (\ref{CF7}) and (\ref{CF8}) to the action
(\ref{CF28}). Thus first we go to the euclidean world, so that $D_0$
becomes real and bounded from below. Next we can use (\ref{CF8}), since the
$q_0$ dependence in $D_0$ does not prevent eq. (\ref{CF8}) from holding,
and
we expand diagrammatically the average value of the (euclidean) action
$S_{\rm eff}$ using $H_0^{(J)}$ as model Hamiltonian.

The diagrams so obtained have the same topological structure as the
standard FHNC diagrams previously discussed, provided the time variable is
added for the external points. Thus these diagrams can also be summed up by
means of the usual FHNC technology, but now the constant $\Delta$
is no longer irrelevant: in fact in one energy denominator (namely the one
of the meson) it is compared with a finite quantity (the meson mass). Thus
$\Delta$ can be regarded as an extra variational parameter.

A more realistic {\sl variational} Hamiltonian than $H_0^{(J)}$ should be used
in connection with effective Hamiltonians of the type discussed above. The
interaction term in $H_0^{(J)}$ should be modified to include
state--dependent correlation effects present in a realistic
$\hat{G}$\cite{FHNC3}.
Moreover, a time (or frequency) dependence could be introduced
in $\lambda$, which would imply adding a time dependence in the internal
points of the FHNC diagrams, or in other words to consider time dependent
correlations. Work in this direction is in progress.

\newpage
{\large\bf Figure Caption}
\begin{enumerate}
\item
\begin{description}
\item{1a} A first order (with respect to $\lambda$) Feynman diagram.
\item{1b} The diagrams obtained from 1a by means of eq. (\ref{CF12}).
\item{1c} The corresponding FHNC diagrams.
\end{description}
\item The ladder series corresponding to eq. (\ref{CF26})
\end{enumerate}


\begin{thebibliography} {99}
\bibitem{FHNC1} S. Fantoni, S. Rosati, Nuovo Cimento {\bf 20A} (1974) 179;
{\bf 25A} (1975) 593;
\bibitem{FHNC2} V. R. Pandharipande, R. B. Wiringa, Rev. Mod. Phys. {\bf
51} (1979) 821;
\bibitem{FHNC3} S. Rosati, in {\sl Nuclei to Particles}, Proceedings of the
International School of Physics "Enrico Fermi", Course LXXIX, edited by A.
Molinari (north--Holland, Amsterdam, 1982), p. 73;
\bibitem{CBF1} J. W. Clark, L. R. Mead, E. Krotscheck, K. E. Kurten, M. L.
Ristig, Nucl. Phys. {\bf A238} (1979) 45;
\bibitem{CBF2} S. Fantoni, V. R. Pandharipande, Phys. Rev. {\bf C37} (1988)
1697;
\bibitem{CBF3} O. Benhar, A. Fabrocini, S. Fantoni, in {\sl Modern Topics
in Electron Scattering}, edited by B. Frois and I. Sick (World Scientific,
Singapore, 1991), p. 460--509;
\bibitem{FHNC6} R. P. Feynman and A. R. Hibbs, ``Quantum Mechanics and Path
Integrals'', McGraw-Hills, N.Y. 1965.
\bibitem{FHNC9} R. Cenni, E. Galleani d'Agliano, F. Napoli, P. Saracco
eand M. Sassetti, ``Feynman Integrals in Theoretical, Nuclear and
Statistical Physics'', ed. Bibliopolis, Napoli 1989.
\bibitem{FHNC7} M. Gari and H. Hyuga, Z. Physik {\bf A227} (1976) 291.
\bibitem{FHNC8} D. O. Riska, in ``Mesons in Nuclei'', vol II, p. 757,
ed. M. Rho and D. H. Wilkinson, North-Holland P. C. Amsterdam 1979.
\bibitem{FHNC11} W. M. Alberico, R. Cenni, A. Molinari and P. Saracco,
Ann. of. Phys.  {\bf 174} (1987) 131.
\bibitem{FHNC12} T. D. Schulz, in ``Polarons and Excitons'', p. 71,
ed. C. G. Kuper
and G. D. Whitfield, Plenum Press, N.Y. 1962.
\bibitem{FHNC10} J. P. Blaizot and G. Ripka, ``Quantum Theory of Finite
Systems'', MIT Press, Cambridge(Ma) 1989.
\end{thebibliography}
\end{document}